\documentclass[12pt,prd,tightenlines,nofootinbib,showpacs]{revtex4}
\usepackage{bm}
\usepackage{graphics}
\usepackage{epsfig}
\begin{document}
\title{\begin{flushright}{\rm\normalsize SSU-HEP-07/12\\[5mm]}\end{flushright}
GROUND STATE HYPERFINE STRUCTURE\\ OF MUONIC HELIUM ATOM
\footnote{Talk presented at the scientific session-conference of
Nuclear Physics Department RAS "Physics of fundamental
interactions", 25-30 November 2007, ITEP, Moscow}}
\author{A.A.Krutov, A.P.Martynenko\footnote{E-mail: mart@ssu.samara.ru (A.P.Martynenko)}}
\affiliation{Samara State University, Pavlov street 1, 443011,
Samara, Russia}

\begin{abstract}
On the basis of the perturbation theory in the fine structure
constant $\alpha$ and the ratio of the electron to muon masses we
calculate one-loop vacuum polarization and electron vertex
corrections and the nuclear structure corrections to the hyperfine
splitting of the ground state of muonic helium atom $(\mu e
^4_2He)$. We obtain total result for the ground state hyperfine
splitting $\Delta \nu^{hfs}=4465.526$ MHz which improves the
previous calculation of Lakdawala and Mohr due to the account of new
corrections. The remaining difference between the theoretical result
and experimental value of the hyperfine splitting equal to 0.522 MHz
lies in the range of theoretical error and requires the subsequent
investigation of higher order corrections.
\end{abstract}


\maketitle

\section{Introduction}

Muonic helium atom $(\mu e ^4_2He)$ represents the simple three-body
atomic system. The interaction between magnetic moments of the muon
and electron leads to the hyperfine structure (HFS) of the energy
levels. The investigation of the energy spectrum of this
three-particle bound state is important for the further check of
quantum electrodynamics. Hyperfine splitting of the ground state of
muonic helium was measured many years ago with sufficiently high
accuracy \cite{Orth1,Orth2}:
\begin{equation}
\Delta\nu^{hfs}_{exp}=4465.004(29)~MHz.
\end{equation}
Contrary to the energy levels of two-particle bound states which
were accurately calculated in quantum electrodynamics
\cite{HBES,SY,EGS,SGK,FM,M3,M4}, the hyperfine splitting of the
ground state in muonic helium atom was calculated on the basis of
the perturbation theory (PT) and the variational method with
essentially less accuracy
\cite{HH,EB,LM1,RD,Amusia,RD1,HH1,LM2,Chen,Chen1,AF,KP2000}. Indeed,
the theoretical errors of the results obtained in
Refs.\cite{HH,EB,LM1,RD,Amusia,RD1,HH1,LM2,Chen,Chen1,AF,KP2000} lie
in the interval $0.05\div 1.8$ MHz. The variational method gives the
high numerical accuracy of the calculation what was demonstrated in
Refs.\cite{RD1,Chen,AF,KP2000}. But higher order corrections were
accounted in this approach less precisely. So, for instance, the
theoretical uncertainty 0.05 MHz in Ref.\cite{Chen} is estimated
only from the numerical convergence of the results obtained by the
variational method for the lowest order hyperfine splitting. The
nonvariational calculation of the lowest order contribution to the
hyperfine structure was performed in the hyperspherical harmonic
method in Ref.\cite{Krivec}. But numerous important corrections to
the hyperfine splitting which are necessary for the successful
comparison with the experimental data were not considered in
\cite{Krivec}.

Many theoretical efforts were focused on the calculation of
different corrections to the Fermi energy which is of the fourth
order over the fine structure constant $\alpha$. The first part of
the calculations was devoted to the recoil corrections which contain
the ratio of the electron and muon masses \cite{LM1,LM2}. The second
group was connected with the relativistic and QED effects which
include another small parameter $\alpha$ \cite{HH,EB,Amusia}. The
particle coordinates in the muonic helium atom were taken in
Refs.\cite{HH,EB,LM1,RD,Amusia,HH1,LM2,Chen,Chen1} by the various
ways. So, numerical results for the corrections of different order
over the fine structure constant $\alpha$ and the ratio of the
particle masses obtained in these papers are difficult for the
direct comparison.

The bound particles in muonic helium atom have different masses
$m_e\ll m_\mu\ll m_\alpha$. As a result the muon and
$\alpha$-particle compose the pseudonucleus $(\mu ^4_2He)^+$ and the
muonic helium atom looks as a two-particle system in the first
approximation. The perturbation theory approach to the investigation
of the hyperfine structure of muonic helium based on the
nonrelativistic Schr\"odinger equation was developed previously by
Lakdawala and Mohr in Refs.\cite{LM1,LM2}. Three-particle bound
system $(\mu e ^4_2He)$ is described by the Hamiltonian:
\begin{equation}
H=H_0+\Delta H+\Delta H_{rec},~~~H_0=-\frac{1}{2M_\mu}\nabla^2_\mu-\frac{1}{2M_e}
\nabla^2_e-\frac{2\alpha}{x_\mu}-\frac{\alpha}{x_e},
\end{equation}
\begin{equation}
\Delta H=\frac{\alpha}{x_{\mu e}}-\frac{\alpha}{x_e},~~~\Delta H_{rec}=-\frac{1}{m_\alpha}
{\mathstrut\bm\nabla}_\mu\cdot{\mathstrut\bm\nabla}_e,
\end{equation}
where ${\bf x_\mu}$ and ${\bf x_e}$ are the coordinates of the muon
and electron relative to the helium nucleus,
$M_e=m_em_\alpha/(m_e+m_\alpha)$, $M_\mu=m_\mu
m_\alpha/(m_\mu+m_\alpha)$ are the reduced masses of subsystems $(e
^4_2He)^+$ and $(\mu ^4_2He)^+$ \cite{LM1,LM2}. The hyperfine part
of the Hamiltonian is
\begin{equation}
\Delta H^{hfs}=-\frac{8\pi\alpha}{3m_em_\mu}
\frac{({\mathstrut\bm\sigma}_e{\mathstrut\bm\sigma}_\mu)}{4}\delta({\bf
x_\mu}-{\bf x_e}),
\end{equation}
where ${\mathstrut\bm \sigma}_e$ and ${\mathstrut\bm\sigma}_\mu$ are
the spin matrices of the electron and muon, $\kappa_e$ and
$\kappa_\mu$ are the electron and muon anomalous magnetic moments.
In the initial approximation the wave function of the ground state
has the form \cite{LM1,LM2}:
\begin{equation}
\Psi_0({\bf x_e},{\bf x_\mu})=\psi_e({\bf x_e})\psi_\mu({\bf x_\mu})=\frac{1}{\pi}
(2\alpha^2M_eM_\mu)^{3/2}e^{-2\alpha M_\mu x_\mu}e^{-\alpha M_e x_e}.
\end{equation}
Then the basic contribution to the singlet-triplet hyperfine splitting can be calculated
analytically from the contact interaction (4):
\begin{equation}
\Delta\nu^{hfs}_0=<\frac{8\pi\alpha}{3m_em_\mu} \delta({\bf
x_\mu}-{\bf
x_e})>=\frac{\nu_F}{\left(1+\frac{M_e}{2M_\mu}\right)^3}, ~~~\nu_F=
\frac{8\alpha^4M_e^3}{3m_em_\mu}.
\end{equation}
Numerically the Fermi splitting is equal $\nu_F=4516.915$ MHz. We
express further the hyperfine splitting contributions in the
frequency unit using the relation $\Delta E^{hfs}=2\pi\hbar\Delta
\nu^{hfs}$. The recoil correction determined by the ratio
$M_e/M_\mu$ in Eq.(6) amounts $\Delta\nu^{hfs}_{rec}=-33.525$ MHz
\cite{LM1}. Modern numerical values of fundamental physical
constants are taken from the paper \cite{MT}: the electron mass
$m_e=0.510998918(44)\cdot 10^{-3}~GeV$, the muon mass
$m_\mu=0.1056583692(94)~GeV$, the fine structure constant
$\alpha^{-1}=137.03599911(46)$, the helium mass $m(^4_2He)$ =
3.72737904(15)~GeV, the electron anomalous magnetic moment
$\kappa_e= 1.1596521869(41)\cdot 10^{-3}$, the muon anomalous
magnetic moment $\kappa_\mu=1.16591981(62)\cdot 10^{-3}$.

Analytical and numerical calculation of corrections which are
determined by the Hamiltonians $\Delta H$ and $\Delta H_{rec}$ in
the second order perturbation theory was performed in
Refs.\cite{LM1,LM2}. Their results and the order of the calculated
contributions are presented in Table I. In this work we aim to
refine the calculation of Lakdawala and Mohr using their approach to
the description of the muonic helium atom. A feature that
distinguishes light muonic atoms among the simplest atoms is that
the structure of their energy levels depends strongly on the vacuum
polarization, nuclear structure and recoil effects
\cite{HBES,SY,EGS,SGK,FM,M3,M4}. So, we investigate such
contributions of the one-loop electron vacuum polarization of order
$\alpha^5M_e/M_\mu$ and the nuclear structure of order $\alpha^6$
which are significant for the improvement of the theoretical value
of the hyperfine splitting. Another purpose of our study consists in
the improved calculation of the electron one-loop vertex corrections
to HFS of order $\alpha^5$ using the analytical expressions of the
Dirac and Pauli form factors of the electron.

\begin{figure}[htbp]
\centering
\includegraphics{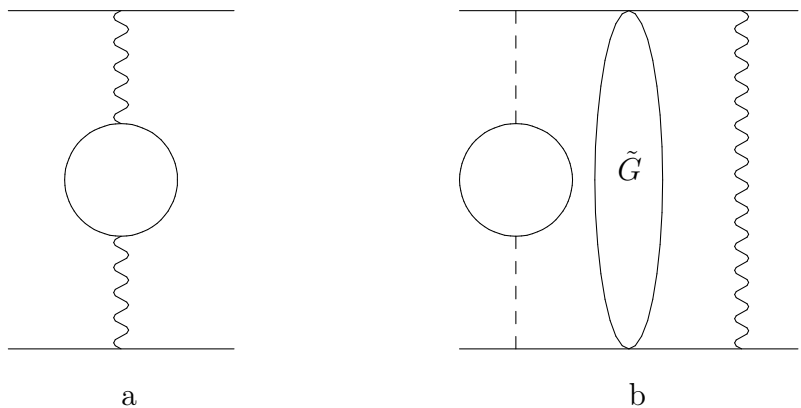}
\caption{The vacuum polarization effects. The dashed line represents
the Coulomb photon. The wave line represents the hyperfine part of
the Breit potential. $\tilde G$ is the reduced Coulomb Green's
function.}
\end{figure}

\section{Effects of the vacuum polarization}

The vacuum effects change the interaction (2)-(3) between particles
in muonic helium atom. One of the most important contributions is
determined by the one-loop vacuum polarization (VP) and electron
vertex operator. The electron vacuum polarization and vertex
corrections to the hyperfine splitting of the ground state contain
the parameter equal to the ratio of the Compton wave length of the
electron and the radius of the Bohr orbit in the subsystem $(\mu
^4_2He)^+$: $m_\mu\alpha/m_e$= $1.50886\ldots$ . The expansion over
$\alpha$ for such contributions to the energy spectrum is unfit to
use. So, we calculate them performing the analytical or numerical
integration over the particle coordinates and other parameters
without an expansion in $\alpha$. The effect of the electron vacuum
polarization leads to the appearance of a number of additional terms
in the interaction operator which we present in the form
\cite{t4,EGS}:
\begin{equation}
\Delta V_{VP}^{e\alpha}(x_e)=\frac{\alpha}{3\pi}\int_1^\infty
\rho(\xi)\left(-\frac{2\alpha}{x_e}\right) e^{-2m_e\xi
x_e}d\xi,~~~\rho(\xi)=\frac{\sqrt{\xi^2-1}(2\xi^2+1)}{\xi^4},
\end{equation}
\begin{equation}
\Delta V_{VP}^{\mu \alpha}(x_\mu)=\frac{\alpha}{3\pi}\int_1^\infty
\rho(\xi)\left(-\frac{2\alpha}{x_\mu}\right) e^{-2m_e\xi x_\mu}d\xi,
\end{equation}
\begin{equation}
\Delta V_{VP}^{e\mu}(|{\bf x}_e-{\bf
x}_\mu|)=\frac{\alpha}{3\pi}\int_1^\infty
\rho(\xi)\frac{\alpha}{x_{e\mu}} e^{-2m_e\xi x_{e\mu}}d\xi,
\end{equation}
where $x_{e\mu}=|{\bf x}_e-{\bf x}_\mu|$. They give contributions to
the hyperfine splitting in the second order perturbation theory and
are discussed below. In the first order perturbation theory the
contribution of the vacuum polarization is connected with the
modification of the hyperfine splitting part of the Hamiltonian (4)
(the diagram (a) in Fig.1). In the coordinate representation it is
determined by the integral expression \cite{KP1996,M1,M2}:
\begin{equation}
\Delta V_{VP}^{hfs}({\bf x}_{e\mu})=-\frac{8\alpha}{3m_em_\mu}
\frac{({\mathstrut\bm\sigma}_1{\mathstrut\bm\sigma}_2)}{4}\frac{\alpha}{3\pi}\int_1^\infty
\rho(\xi)d\xi\left[\pi\delta({\bf
x_{e\mu}})-\frac{m_e^2\xi^2}{x_{e\mu}}e^{-2m_e\xi x_{e\mu}} \right].
\end{equation}
Averaging the potential (10) over the wave function (5) we obtain
the following contribution to the hyperfine splitting:
\begin{equation}
\Delta\nu^{hfs}_{VP}=\frac{8\alpha^2}{9m_em_\mu}\frac{(\alpha
M_e)^3(2\alpha M_\mu)^3} {\pi^3}\int_1^\infty\rho(\xi)d\xi\int d{\bf
x}_e\int d{\bf x}_\mu e^{-4\alpha M_\mu x_\mu}e^{-2\alpha
M_ex_e}\times
\end{equation}
\begin{displaymath}
\times\left[\pi\delta({\bf x_\mu}-{\bf x}_e)-\frac{m_e^2\xi^2}{|{\bf x}_\mu-{\bf x}_e|}\right]
e^{-2m_e\xi|{\bf x}_\mu-{\bf x}_e|}.
\end{displaymath}
There are two integrals over the muon and electron coordinates in
Eq.(11) which can be calculated analytically:
\begin{equation}
I_1=\int d{\bf x}_e\int d{\bf x}_\mu e^{-4\alpha M_\mu x_\mu}e^{-2\alpha M_ex_e}\pi
\delta({\bf x_\mu}-{\bf x}_e)
=\frac{\pi^2}{8\alpha^3M_\mu^3\left(1+\frac{M_e}{2M_\mu}\right)^3},
\end{equation}
\begin{equation}
I_2=\int d{\bf x}_e\int d{\bf x}_\mu e^{-4\alpha M_\mu x_\mu}e^{-2\alpha M_ex_e}\frac{1}
{|{\bf x}_\mu-{\bf x}_e|}e^{-2m_e\xi|{\bf x}_\mu-{\bf x}_e|}=
\end{equation}
\begin{displaymath}
=\frac{32\pi^2}{(4\alpha
M_\mu)^5}\frac{\left[\frac{M_e^2}{4M_\mu^2}+\left(1+
\frac{m_e\xi}{2M_\mu\alpha}\right)^2+\frac{M_e}{2M_\mu}\left(3+\frac{m_e\xi}{M_\mu
\alpha}\right)\right]}{\left(1+\frac{M_e}{2M_\mu}\right)^3\left(1+\frac{m_e\xi}
{2M_\mu\alpha}\right)^2\left(\frac{M_e}{2M_\mu}+\frac{m_e\xi}{2M_\mu\alpha}\right)^2}.
\end{displaymath}
They are divergent separately in the subsequent integration over the
parameter $\xi$. But their sum is finite and can be written in the
integral form:
\begin{equation}
\Delta\nu^{hfs}_{VP}=\nu_F\frac{\alpha M_e}{6\pi
M_\mu\left(1+\frac{M_e}{2M_\mu}\right)^3}\int_1^\infty\rho(\xi)d\xi
\frac{\left[\frac{M_e}{2M_\mu}+2\frac{m_e\xi}{2M_\mu\alpha}\frac{M_e}{2M_\mu}+
\frac{m_e\xi}{2M_\mu\alpha}\left(2+\frac{m_e\xi}{2M_\mu
\alpha}\right)\right]}{\left(1+\frac{m_e\xi}
{2M_\mu\alpha}\right)^2\left(\frac{M_e}{2M_\mu}+\frac{m_e\xi}{2M_\mu\alpha}\right)^2}=0.035~MHz.
\end{equation}
The order of this contribution is determined by two small parameters
$\alpha$ and $M_e/M_\mu$ which are written explicitly. The
correction $\Delta\nu_{VP}^{hfs}$ is of the fifth order in $\alpha$
and the first order in the ratio of the electron and muon masses.
The contribution of the muon vacuum polarization to the hyperfine
splitting is extremely small ($\sim 10^{-6}$ MHz). One should expect
that two-loop vacuum polarization contributions to the hyperfine
structure are suppressed relative to the one-loop VP contribution by
the factor $\alpha/\pi$. This means that at present level of
accuracy we can neglect these corrections because their numerical
value is not exceeding 0.001 MHz. Higher orders of the perturbation
theory which contain one-loop vacuum polarization and the Coulomb
interaction (3) lead to the recoil corrections of order
$\nu_F\alpha\frac{M_e^2}{M_\mu^2}\ln\frac{M_\mu}{M_e}$. Such terms
which can contribute 0.004 MHz are included in the theoretical
error.

It is useful to compare the obtained result (14) with the
calculation of VP contribution to the HFS in which the expansion in
$\alpha$ is used. Instead of the potential (10) we obtain the
following operator in the coordinate representation:
\begin{equation}
\Delta\tilde V^{hfs}_{VP}({\bf
x}_{e\mu})=-\frac{8\pi\alpha}{3m_em_\mu}\frac{\alpha}{15\pi}\frac{1}{m_e^2}
\nabla^2\delta({\bf x}_{e\mu}).
\end{equation}
Then the contribution of (15) to the hyperfine structure can be
derived in the analytical form:
\begin{equation}
\Delta \tilde
\nu_{VP}^{hfs}=\nu_F\frac{8\alpha}{15\pi}\left(\frac{\alpha
M_\mu}{m_e}\right)^2\frac{M_e}{M_\mu}\frac{1}{\left(1+\frac{M_e}{2M_\mu}\right)^3}=0.060~MHz.
\end{equation}
This calculation demonstrates the need to employ the exact
potentials (7)-(9) for the study of the electron vacuum polarization
corrections.

Let us consider corrections of the electron vacuum polarization
(7)-(9) in the second order perturbation theory (SOPT) (the diagram
(b) in Fig.1). The contribution of the electron-nucleus interaction
(7) to the hyperfine splitting can be written as follows:
\begin{equation}
\Delta\nu_{VP~SOPT~e\alpha}^{hfs}=\frac{16\pi\alpha}{3m_em_\mu}\int
d{\bf x}_1\int d{\bf x}_2\int d{\bf x}_3
\frac{\alpha}{3\pi}\int_1^\infty\rho(\xi)d\xi\psi^\ast_{\mu 0}({\bf
x}_3)\psi^\ast_{e 0}({\bf x}_3)\times
\end{equation}
\begin{displaymath}
\times\sum_{n,n'\not =0}^\infty\frac{\psi_{\mu n}({\bf x}_3) \psi_{e
n'}({\bf x}_3)\psi^\ast_{\mu n}({\bf x}_2)\psi^\ast_{e n'}({\bf
x}_1)}{E_{\mu 0}+E_{e0}-E_{\mu n}-E_{en'}}e^{-2m_e\xi x_1}\psi_{\mu
0}({\bf x}_2) \psi_{e0}({\bf x}_1).
\end{displaymath}
Here the summation is carried out over the complete system of the
eigenstates of the electron and muon excluding the state with
$n,n'=0$. The computation of the expression (17) is simplified with
the use of the orthogonality condition for the muon wave functions:
\begin{equation}
\Delta\nu_{VP~SOPT~e\alpha}^{hfs}=\nu_F\frac{32\alpha M_e^2}{3\pi
M_\mu^2}\int_1^\infty \rho(\xi)d\xi\int _0^\infty x_3^3 dx_3\int
_0^\infty x_1 dx_1
e^{-x_1\frac{M_e}{M_\mu}\left(1+\frac{m_e\xi}{\alpha M_\mu}\right)}
e^{-2x_3\left(1+\frac{M_e}{2M_\mu}\right)}\times
\end{equation}
\begin{displaymath}
\left[\frac{M_\mu}{M_ex_>}-\ln\left(\frac{M_e}{M_\mu}x_<\right)-\ln
\left(\frac{M_e}{M_\mu}x_>\right)+Ei\left(\frac{M_e}{M_\mu}x_<\right)+
\frac{7}{2}-2C-\frac{M_e}{2M_\mu}(x_1+x_3)+
\frac{1-e^{\frac{M_e}{M_\mu}x_<}}{\frac{M_e}{M_\mu}x_<}\right]=
\end{displaymath}
\begin{displaymath}
=0.151~MHz,
\end{displaymath}
where $x_<=\min(x_1,x_3)$, $x_>=\max(x_1,x_3)$, $C=0.577216\ldots$
is the Euler's constant and $Ei(x)$ is the exponential-integral
function. It is necessary to emphasize that the transformation of
the expression (17) into (18) is carried out with the use of the
compact representation for the electron reduced Coulomb Green's
function obtained in Refs.\cite{Hameka,LM1}:
\begin{equation}
G_e({\bf x}_1,{\bf x}_3)=\sum_{n\not =0}^\infty\frac{\psi_{en}({\bf
x}_3) \psi_{en}^\ast({\bf x}_1)}{E_{e0}-E_{en}}=-\frac{\alpha
M_e^2}{\pi}e^{-\alpha M_e(x_1+x_3)}\Biggl[\frac{1}{2\alpha M_e x_>}-
\end{equation}
\begin{displaymath}
-\ln(2\alpha M_e x_>)-\ln(2\alpha M_e x_<)+Ei(2\alpha M_e x_<)+
\frac{7}{2}-2C-\alpha M_e(x_1+x_3)+\frac{1-e^{2\alpha M_e
x_<}}{2\alpha M_e x_<}\Biggr].
\end{displaymath}
The contribution (18) has the same order of the magnitude
$O(\alpha^5\frac{M_e}{M_\mu})$ as the previous correction (14) in
the first order perturbation theory. Similar calculation can be
performed in the case of muon-nucleus vacuum polarization operator
(8). The intermediate electron state is the 1S state and the reduced
Coulomb Green's function of the system appearing in the second order
PT transforms to the Green's function of the muon. The correction of
the operator (8) to the hyperfine splitting is obtained in the
following integral form:
\begin{equation}
\Delta\nu^{hfs}_{VP~SOPT~\mu
\alpha}=\nu_F\frac{\alpha}{3\pi}\int_1^\infty\rho(\xi)d\xi\int_0^\infty
x_3^2dx_3\int_0^\infty x_2dx_2
e^{-x_3\left(1+\frac{M_e}{2M_\mu}\right)}e^{-x_2\left(1+\frac{m_e\xi}
{2M_\mu\alpha}\right)}\times
\end{equation}
\begin{displaymath}
\times\left[\frac{1}{x_>}-\ln x_>-\ln x_<+Ei (x_<)+\frac{7}{2}-2C
-\frac{x_2+x_3}{2}+\frac{1-e^{x_<}}{x_<}\right]= 0.048~MHz.
\end{displaymath}
The vacuum polarization correction to HFS which is determined by the
operator (9) in the second order perturbation theory is the most
difficult for the calculation. Indeed, in this case we have to
consider the intermediate excited states both for the muon and
electron. Following Ref.\cite{LM1} we have divided total
contribution into two parts. The first part in which the
intermediate muon is in the 1S state can be written as:
\begin{equation}
\Delta\nu^{hfs}_{VP~SOPT~\mu e}(n=0)=\frac{256\alpha^2(\alpha
M_e)^3(2\alpha M_\mu)^3}{9} \int_0^\infty x_3^2dx_3\times
\end{equation}
\begin{displaymath}
\times\int_0^\infty x_1^2 dx_1 e^{-\alpha(M_e+4M_\mu)x_3}\int_1^\infty\rho(\xi)
d\xi\Delta V_{VP~\mu}(x_1)G_e(x_1,x_3),
\end{displaymath}
where the function $V_{VP~\mu}(x_1)$ is equal
\begin{equation}
\Delta V_{VP~\mu}(x_1)=\int d{\bf x}_2 e^{-4\alpha M_\mu
x_2}\frac{(2\alpha M_\mu)^3} {\pi}\frac{\alpha}{|{\bf x}_1-{\bf
x}_2|}e^{-2m_e\xi|{\bf x}_1-{\bf x}_2|}=
\end{equation}
\begin{displaymath}
=\frac{32\alpha^4 M^3_\mu}{x_1(16\alpha^2 M_\mu^2-4m_e^2\xi^2)^2}
\left[8\alpha M_\mu\left(e^{-2m_e\xi x_1}-e^{-4\alpha M_\mu
x_1}\right)+x_1(4m_e^2\xi^2-16\alpha^2 M_\mu^2)e^{-4\alpha M_\mu
x_1}\right].
\end{displaymath}
After the substitution (22) in (21) the numerical integration gives
the result:
\begin{equation}
\Delta\nu^{hfs}_{VP~SOPT~\mu e}(n=0)=-0.030~MHz.
\end{equation}
Second part of the vacuum polarization correction to the hyperfine
splitting due to the electron-muon interaction (9) can be presented
as follows:
\begin{equation}
\Delta\nu^{hfs}_{VP~SOPT~\mu e}(n\not
=0)=-\frac{16\alpha^2}{9m_em_\mu}\int d{\bf x}_3\int d{\bf
x}_2\int_1^\infty\rho(\xi)d\xi \psi^\ast_{\mu 0}({\bf
x}_3)\psi^\ast_{e 0}({\bf x}_3)\times
\end{equation}
\begin{displaymath}
\times\sum_{n\not =0}\psi_{\mu n}({\bf x}_3)\psi^\ast_{\mu n}({\bf
x}_2) \frac{M_e}{2\pi}\frac{e^{-b|{\bf x}_3-{\bf x}_1|}}{|{\bf
x}_3-{\bf x}_1|}\frac{\alpha}{|{\bf x}_2-{\bf x}_1|}e^{-2m_e\xi|{\bf
x}_2-{\bf x}_1|} \psi_{\mu 0}({\bf x}_2)\psi_{e 0}({\bf x}_1).
\end{displaymath}
In the expression (24) we have replaced the exact electron Coulomb
Green's function by the free electron Green's function which
contains $b=[2M_e(E_{\mu n}-E_{\mu 0}-E_{e 0}]^{1/2}$. (see more
detailed discussion of this approximation in Refs.\cite{LM1,LM2}).
We also replace the electron wave functions by their values at the
origin as in Ref.\cite{LM1} neglecting higher order recoil
corrections. After that the integration over ${\bf x}_1$ can be done
analytically:
\begin{equation}
J=\int d{\bf x}_1\frac{e^{-b|{\bf x}_3-{\bf x}_1|}}{|{\bf x}_3-{\bf
x}_1|} \frac{e^{-2m_e\xi|{\bf x}_2-{\bf x}_1|}}{|{\bf x}_2-{\bf
x}_1|}= -\frac{4\pi}{|{\bf x}_3-{\bf
x}_2|}\frac{1}{b^2-4m_e^2\xi^2}\left[e^{-b|{\bf x}_3-{\bf
x}_2|}-e^{-2m_e\xi|{\bf x}_3-{\bf x}_2|}\right]=
\end{equation}
\begin{displaymath}
=2\pi\Biggl[\frac{\left(1-e^{-2m_e\xi|{\bf x}_3-{\bf x}_2|}\right)}
{2m_e^2\xi^2|{\bf x}_3-{\bf x}_2|}-
\frac{b}{2m_e^2\xi^2}+\frac{\left(1-e^{-2m_e\xi|{\bf x}_3-{\bf x}_2|}\right)b^2}
{8m_e^4\xi^4|{\bf x}_3-{\bf x}_2|}+\frac{b^2|{\bf x}_3-{\bf x}_2|}{4m_e^2\xi^2}-
\end{displaymath}
\begin{displaymath}
-\frac{b^3}{8m_e^4\xi^4}-\frac{b^3({\bf x}_3-{\bf x}_1)^2}{12m_e^2\xi^2}+...\Biggr],
\end{displaymath}
where we have performed the expansion of the first exponential in
the square brackets over powers of $b|{\bf x}_3-{\bf x}_2|$. As
discussed in Ref.\cite{LM1} one can treat this series as an
expansion over the recoil parameter $\sqrt{M_e/M_\mu}$. For the
further transformation the completeness condition is useful:
\begin{equation}
\sum_{n\not =0}\psi_{\mu n}({\bf x}_3)\psi_{\mu n}^\ast({\bf x}_2)=
\delta({\bf x}_3-{\bf x}_2)-\psi_{\mu 0}({\bf x}_3)\psi_{\mu
0}^\ast({\bf x}_2).
\end{equation}
The wave function orthogonality leads to the zero results for the
second and fifth terms in the square brackets of Eq.(25). The first
term in Eq.(25) gives the leading order contribution in two small
parameters $\alpha$ and $M_e/M_\mu$:
\begin{equation}
\Delta\nu^{hfs}_{VP~SOPT~\mu e}(n\not =0)=\Delta\nu_{11}+\Delta
\nu_{12},~\Delta\nu_{11}= -\frac{3\alpha^2M_e}{8m_e}\nu_F,
\end{equation}
\begin{equation}
\Delta\nu_{12}=\nu_F\frac{2\alpha^2}{3\pi\frac{m_e}{M_e}}\int_1^\infty\rho(\xi)
\frac{d\xi}{\xi}\frac{M_\mu^4\alpha^4}{(4\alpha
M_\mu+2m_e\xi)^4}\left[256+232\frac{m_e\xi}{M_\mu \alpha}+80\frac{
m_e^2\xi^2}{M_\mu^2\alpha^2}+
10\frac{m_e^3\xi^3}{M_\mu^3\alpha^3}\right].
\end{equation}
The numerical value of the sum $\Delta\nu_{11}+\Delta\nu_{12}$ is
included in Table I. It is important to calculate also the
contributions of other terms of the expression (25) to the hyperfine
splitting. Taking the fourth term in Eq.(25) which is proportional
to $b^2=2M_e(E_{\mu n}-E_{\mu 0})$ we have performed the sequence of
the transformations in the coordinate representation:
\begin{equation}
\sum_{n=0}^\infty E_{\mu n}\int d{\bf x}_2\int d{\bf x}_3\psi_{\mu
0}^\ast({\bf x}_2) \psi_{\mu n}({\bf x}_3)\psi_{\mu n}^\ast({\bf
x}_2)|{\bf x}_3-{\bf x}_2|\psi_{\mu 0}({\bf x}_2)=
\end{equation}
\begin{displaymath}
=\int d{\bf x}_2\int d{\bf x}_3\delta({\bf x}_3-{\bf x}_2)
\left[-\frac{\nabla^2_3}{2M_\mu}|{\bf x}_3-{\bf x}_2|\psi_{\mu 0}^\ast({\bf x}_3)\right]
\psi_{\mu 0}({\bf x}_2).
\end{displaymath}
Evidently, we have the divergent expression in Eq.(29) due to the
presence of the $\delta$-function. The same divergence occurs in the
other term containing $b^2$ in the square brackets of Eq.(25). But
their sum is finite and can be calculated analytically with the
result:
\begin{equation}
\Delta\nu^{hfs}_{b^2}=\nu_F\frac{\alpha^2
M_e^2}{m_eM_\mu}\left(18-5\frac{\alpha^2 M_\mu^2}{m_e^2}\right).
\end{equation}
Numerical value of this correction 0.0002 MHz is essentially smaller
than the leading order term. Let us consider also the nonzero term
in Eq.(25) proportional to $b^3$. First of all, it can be
transformed to the following expression after the integration over
$\xi$:
\begin{equation}
\Delta\nu^{hfs}_{b^3}=-\nu_F\frac{4\alpha^3}{45\pi}\sqrt{\frac{M_e}{M_\mu}}
\frac{M_e^2}{m_e^2}S_{3/2},
\end{equation}
where the sum $S_{3/2}$ is defined as follows:
\begin{equation}
S_p=\sum_n\left[\left(\frac{E_{\mu n}-E_{\mu
0}}{R_\mu}\right)^{p}\right] |<\psi_{\mu 0}|\frac{{\bf
x}}{a_\mu}|\psi_{\mu n}>|^2,
\end{equation}
$R_\mu=2\alpha^2 M_\mu$, $a_\mu=1/2\alpha M_\mu$. Using the known
analytical expressions for the dipole matrix elements entering in
Eq.(32) in the case of the discrete and continuous spectrum
\cite{HBES,VAF} we can write separately their contributions to the
sum $S_{3/2}$ in the form:
\begin{equation}
S_{3/2}^d=\sum_{n=0}^\infty\frac{2^8n^4(n-1)^{2n-\frac{7}{2}}}{(n+1)^{2n+
\frac{7}{2}}}=1.50989\ldots,
\end{equation}
\begin{equation}
S_{3/2}^c=\int_0^\infty k dk \frac{2^8}{\left(1-e^{-\frac{2\pi}{k}}\right)}
\frac{1}{(1+k^2)^{\frac{7}{2}}}
|\left(\frac{1+ik}{1-ik}\right)^{\frac{i}{k}}|^2=1.76236\ldots  .
\end{equation}
As a result $S_{3/2}=3.2722\ldots$ . The similar calculation of the
sum $S_{1/2}$ relating to this problem (see Ref.\cite{LM1}) gives
$S_{1/2}=2.9380\ldots$. Numerical value (31) is taken into account
in the total result presented in Table I.

\begin{figure}[htbp]
\centering
\includegraphics{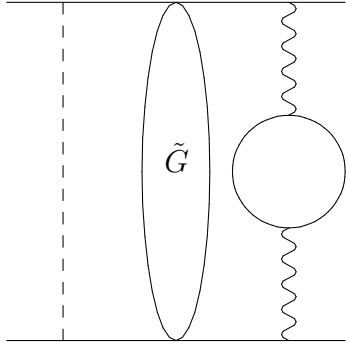}
\caption{Vacuum polarization effects in the second order
perturbation theory. The dashed line represents the first part of
the potential $\Delta H$ (3). The wave line represents the hyperfine
part of the Breit potential.}
\end{figure}

There exists another contribution of the second order perturbation
theory in which we have the vacuum polarization perturbation
connected with the hyperfine splitting part of the Breit potential
(10) (see Fig.2). Another perturbation potential in this case is
determined by the first term of relation (3). We can divide this
correction into two parts as previously. One part with $n=0$
corresponds to the ground state muon. The other part with $n\not =0$
accounts the excited muon states. The $\delta$-function term in
Eq.(10) gives the following contribution to HFS at $n=0$ (compare
with Ref.\cite{LM1}):
\begin{equation}
\Delta\nu^{hfs}_{VP~SOPT~11}(n=0)=\nu_F\frac{\alpha}{3\pi}\int_1^\infty\rho(\xi)d\xi
\frac{11M_e}{16 M_\mu}.
\end{equation}
Obviously, this integral in the variable $\xi$ is divergent. So, we
have to consider the contribution of the second term of the
potential (10) to the hyperfine splitting which is determined by the
more complicated expression:
\begin{equation}
\Delta\nu^{hfs}_{VP~SOPT~12}(n=0)=\frac{16\alpha^2m_e^2}{9\pi
m_em_\mu}\int_1^\infty \rho(\xi)\xi^2d\xi\int d{\bf x}_3\psi_{e
0}({\bf x}_3)\Delta V_1({\bf x}_3)\times
\end{equation}
\begin{displaymath}
\times\sum_{n'\not =0}
\frac{\psi_{e n'}({\bf x}_3)\psi^\ast_{e n'}({\bf x}_1)}{E_{e0}-E_{en'}}\Delta V_2({\bf x}_1)
\psi_{e 0}({\bf x}_1),
\end{displaymath}
where
\begin{equation}
\Delta V_1({\bf x}_3)=\int d{\bf x}_4\psi^\ast_{\mu 0}({\bf x}_4)
\frac{e^{-2m_e\xi|{\bf x}_3-{\bf x}_4|}}
{|{\bf x}_3-{\bf x}_4|}\psi_{\mu 0}({\bf x}_4)=
\end{equation}
\begin{displaymath}
=\frac{4(2\alpha M_\mu)^3}{x_3[(4\alpha M_\mu)^2-(2m_e\xi)^2]^2}\left[8\alpha M_\mu e^{-2m_e\xi x_3}+
e^{-4\alpha M_\mu x_3}\left(-8\alpha M_\mu-16\alpha^2 M_\mu^2 x_3+4m_e^2\xi^2 x_3\right)\right],
\end{displaymath}
\begin{equation}
\Delta V_2({\bf x}_1)=\int d{\bf x}_2\psi_{\mu 0}({\bf
x}_2)\left(\frac{\alpha}{|{\bf x}_2-{\bf x}_1|}-
\frac{\alpha}{x_1}\right)\psi_{\mu 0}({\bf x}_2)
=-\frac{\alpha}{x_1}(1+2\alpha M_\mu x_1)e^{-4\alpha M_\mu x_1}.
\end{equation}
Nevertheless, integrating over all coordinates in Eq(36) we obtain
the following result in the leading order with respect to the ratio
$(M_e/M_\mu)$:
\begin{equation}
\Delta\nu^{hfs}_{VP~SOPT~12}(n=0)=\nu_F\frac{m_e}{M_e}\frac{M_e^2}{96\pi
M_\mu^2}\int_1^\infty\rho(\xi)\xi
d\xi\frac{32+63\gamma+44\gamma^2+11\gamma^3}{(1+\gamma)^4},
\end{equation}
where $\gamma=m_e\xi/2\alpha M_\mu$. This integral also has the
divergence at large values of the parameter $\xi$. But the sum of
integrals (35) and (39) is finite:
\begin{equation}
\Delta\nu^{hfs}_{VP~SOPT~11}(n=0)+\Delta
\nu^{hfs}_{VP~SOPT~12}(n=0)=0.008~MHz.
\end{equation}

\begin{figure}[htbp]
\centering
\includegraphics{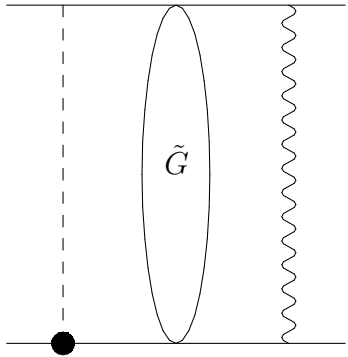}
\caption{Nuclear structure effects in the second order perturbation theory. The bold
point represents the nuclear vertex operator.
The wave line represents the hyperfine part of the Breit potential.}
\end{figure}

Let us consider now the terms with $n\not =0$. The delta-like term
of the potential (10) gives the contribution to HFS known from the
calculation of Ref.\cite{LM1}:
\begin{equation}
\Delta\nu^{hfs}_{VP~SOPT~21}(n\not
=0)=\nu_F\frac{\alpha}{3\pi}\int_1^\infty\rho(\xi)d\xi\left(-\frac{35M_e}{16M_\mu}\right).
\end{equation}
Another correction from the second term of the expression (10) can
be simplified after the replacement the exact electron Green's
function by the free electron Green's function:
\begin{equation}
\Delta\nu^{hfs}_{VP~SOPT~22}(n\not
=0)=-\frac{16\alpha^3M_em_e^2}{9m_em_\mu}
\int_1^\infty\rho(\xi)\xi^2d\xi\int d{\bf x}_2\int d{\bf x}_3\times
\end{equation}
\begin{displaymath}
\times\int d{\bf x}_4\psi_{\mu 0}^\ast({\bf x}_4)
\frac{e^{-2m_e\xi|{\bf x}_3-{\bf x}_4|}}{|{\bf x}_3-{\bf x}_4|}\sum_{n\not=0}^\infty
\psi_{\mu n}({\bf x}_4)\psi_{\mu n}({\bf x}_2)|{\bf x}_3-{\bf x}_2|\psi_{\mu 0}({\bf x}_2)
\end{displaymath}
The analytical integration in Eq.(42) over all coordinates leads to
the result:
\begin{equation}
\Delta\nu^{hfs}_{VP~SOPT~22}(n\not =0)=-\nu_F\frac{\alpha M_e}{3\pi
M_\mu}\int_1^\infty\rho(\xi)d\xi\left[
\frac{1}{\gamma}-\frac{1}{(1+\gamma)^4}\left(4+\frac{1}{\gamma}+10\gamma+\frac{215\gamma^2}{16}+
\frac{35\gamma^4}{16}\right)\right].
\end{equation}
The sum of expressions (41) and (43) gives again the finite
contribution to the hyperfine splitting:
\begin{equation}
\Delta\nu^{hfs}_{VP~SOPT~21}(n\not =0)+\Delta
\nu^{hfs}_{VP~SOPT~22}(n\not =0)=
\end{equation}
\begin{displaymath}
=-\nu_F\frac{\alpha M_e}{3\pi M_\mu}
\int_1^\infty\rho(\xi)d\xi\frac{35+76\gamma+59\gamma^2+16\gamma^3}{16(1+\gamma)^4}=-0.062~MHz.
\end{displaymath}
Despite the fact that the absolute values of the calculated VP
corrections (23), (27), (28), (30), (31), (40), (44) are
sufficiently large , their summary contribution to the hyperfine
splitting (see Table I) is small because they have different signs.

\begin{table}
\caption{\label{t1}Hyperfine singlet-triplet splitting of the ground
state in the muonic helium atom.}
\bigskip
\begin{ruledtabular}
\begin{tabular}{|c|c|c|}  \hline
Contribution to the HFS & $\Delta\nu^{hfs}$, MHz & Reference   \\
\hline The Fermi splitting & 4516.915  & (6), \cite{LM1}  \\  \hline
Recoil correction of order  & -33.525 &   (6),\cite{LM1}  \\
$\alpha^4(m_e/m_\mu)$    &       &      \\    \hline Correction of
muon anomalous &  5.244 &  \cite{LM2,EB}  \\
magnetic moment of order $\alpha^5$&  &  \\  \hline
Recoil correction of order &   0.079   & \cite{LM1,LM2}   \\
$\alpha^4(M_e/m_\alpha)\sqrt{(M_e/M_\mu)}$   &   &        \\     \hline
Correction due to the perturbation&  &    \\
(3) in the second order PT & -29.650     &   \cite{LM1,LM2}   \\
of order $\alpha^4\frac{M_e}{M_\mu}$  &   &    \\    \hline
Relativistic correction of order $\alpha^6$  & 0.040  & \cite{HH} \\ \hline
One-loop VP contribution in $1\gamma$ &  0.035 &  (14)     \\
$(e\mu)$-interaction of order $\alpha^5\frac{M_e}{M_\mu}$   &  &     \\
\hline
One-loop VP contribution in the &   &  (23),(27),(28),(30),     \\
electron-muon interaction in the&   -0.145      &   (31),(40),(44)         \\
second order PT of order $\alpha^5\frac{M_e}{M_\mu}$   &     &          \\
\hline
One-loop VP contribution in the &  &      \\
electron-nucleus interaction in the&  0.151       &    (18)        \\
second order PT of order $\alpha^5\frac{M_e}{M_\mu}$   &     &          \\
\hline
One-loop VP contribution in the &   &      \\
muon-nucleus interaction in the&   0.048      &  (20)          \\
second order PT of order $\alpha^5\frac{M_e}{M_\mu}$   &     &          \\
\hline
Nuclear structure correction of order  & -0.010 & (47),(49) \\
$\alpha^6$ in the second order PT &       &        \\   \hline
Recoil correction of order & 0.812  &  (51) ,\cite{Chen} \\
$\alpha^5(m_e/m_\mu)\ln(m_e/m_\mu)$    &     &    \\   \hline Vertex
correction of order $\alpha^6$ &  -0.606 &  \cite{BE,KP,KKS,EGS}  \\
\hline
Electron vertex contribution &  6.138  &  (53),(56),(58),    \\
of order $\alpha^5$ &       &  (59),(61),(62)     \\
\hline
Summary contribution & $4465.526$  &      \\
\hline
\end{tabular}
\end{ruledtabular}
\end{table}

\section{Nuclear structure and recoil effects}

Another significant corrections to the hyperfine splitting of muonic
helium atom which we study in this work are determined by the
nuclear structure effects. They are specific for any muonic atom. In
the leading order over $\alpha$ they are described by the charge
radius of $\alpha$-particle $r_\alpha$. If we consider the
interaction between the muon and the nucleus then the nuclear
structure correction to the interaction operator has the form
\cite{EGS}:
\begin{equation}
\Delta V_{str,\mu}({\bf r}_\mu)=\frac{2}{3}\pi Z\alpha<r^2_\alpha>\delta({\bf r}_\mu).
\end{equation}
The contribution of the operator $\Delta V_{str,\mu}$ to the
hyperfine splitting appears in the second order perturbation theory
(see the diagram in Fig.3). In the beginning we can write it in the
integral form:
\begin{equation}
\Delta\nu_{str,\mu}^{hfs}=\frac{64\pi^2\alpha^2}{9m_em_\mu}r_\alpha^2
\frac{1}{\sqrt{\pi}}\left(2\alpha M_\mu\right)^{3/2} \int d{\bf
x}_3\psi^\ast_{\mu 0}({\bf x}_3)|\psi_{e0}({\bf x}_3)|^2G_\mu({\bf
x}_3,0,E_{\mu 0}).
\end{equation}
After that the analytical integration over the coordinate ${\bf
x}_3$ in Eq.(46) can be carried out using the representation of the
muon Green's function similar to expression (17). The result of the
integration of order $O(\alpha^6)$ is written as an expansion in the
ratio $M_e/M_\mu$:
\begin{equation}
\Delta\nu_{str,\mu}^{hfs}=-\nu_F\frac{8}{3}\alpha^2M_\mu^2
r_\alpha^2\left(3\frac{M_e}{M_\mu}-
\frac{11}{2}\frac{M_e^2}{M_\mu^2}+\ldots\right)=-0.007 ~MHz.
\end{equation}
Numerical value of the contribution $\Delta\nu_{str,\mu}^{hfs}$ is
obtained by means of the charge radius of the $\alpha$-particle
$r_\alpha=1.676$ fm. The same approach can be used in the study of
the electron-nucleus interaction. The electron feels as well the
distribution of the electric charge of $\alpha$ particle. The
corresponding contribution of the nuclear structure effect to the
hyperfine splitting is determined by the expression:
\begin{equation}
\Delta\nu_{str,e}^{hfs}=\frac{64\pi^2\alpha^2}{9m_em_\mu}r_\alpha^2
\int d{\bf x}_1 \int d{\bf x}_3|\psi^\ast_{\mu 0}({\bf
x}_3)|^2\psi_{e0}({\bf x}_3)G_\mu({\bf x}_3,{\bf x}_1,E_{e 0})
\psi_{e0}({\bf x}_1)\delta({\bf x}_1).
\end{equation}
Performing the analytical integration in Eq.(46) we obtain the
following series:
\begin{equation}
\Delta\nu_{str,e}^{hfs}=-\nu_F\frac{4}{3}\alpha^2
M_e^2r_\alpha^2\left[5-\ln\frac{M_e}{M_\mu}+
\frac{M_e^2}{M_\mu^2}\left(3\ln\frac{M_e}{M_\mu}-7\right)+\frac{M_e^2}{M_\mu^2}\left(\frac{17}{2}-
3\ln\frac{M_e}{M_\mu}\right)\ldots\right]=
\end{equation}
\begin{displaymath}
=-0.003~MHz.
\end{displaymath}
We have included in Table I the total nuclear structure contribution
which is equal to the sum of the numerical values (45) and (47).

\begin{figure}[htbp]
\centering
\includegraphics{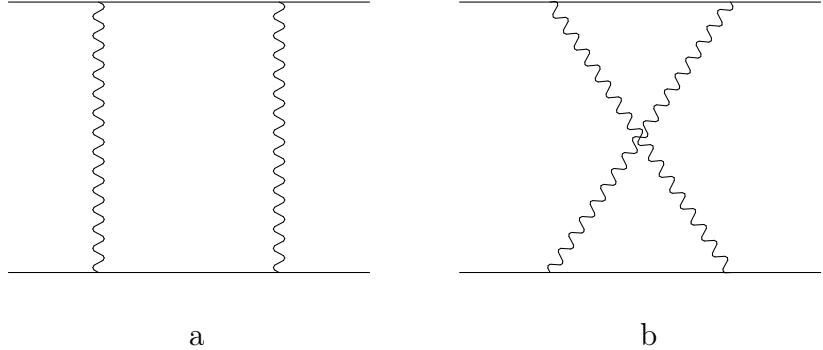}
\caption{Two photon exchange amplitudes in the electron-muon hyperfine
interaction.}
\end{figure}

Special attention has to be given to the recoil corrections
connected with two-photon exchange diagrams shown in Fig.4 in the
case of the electron-muon interaction.  For the singlet-triplet
splitting the leading order recoil contribution to the interaction
operator between the muon and electron is determined as follows
\cite{EGS,Chen,RA}:
\begin{equation}
\Delta V_{rec,\mu e}^{hfs}({\bf x}_{\mu e})=8\frac{\alpha^2}{m_\mu^2-m_e^2}\ln\frac{m_\mu}{m_e}
\delta({\bf x}_{\mu e}).
\end{equation}
Averaging the potential $\Delta V_{rec,\mu e}^{hfs}$ over the wave
functions (5) we obtain the leading order recoil correction to the
hyperfine splitting:
\begin{equation}
\Delta\nu_{rec,\mu
e}^{hfs}=\nu_F\frac{3\alpha}{\pi}\frac{m_em_\mu}{m_\mu^2-m_e^2}\ln\frac{m_\mu}
{m_e}=0.812~MHz.
\end{equation}
There exist also the two-photon interactions between the bound
particles of muonic helium atom when one hyperfine photon transfers
the interaction from the electron to muon and another Coulomb photon
from the electron to the nucleus (or from the muon to the nucleus).
Supposing that these amplitudes give smaller contribution to the
hyperfine splitting we included them in the theoretical error.

\section{Electron vertex corrections}

In the initial approximation the potential of the hyperfine
splitting is determined by Eq.(4). It leads to the energy splitting
of order $\alpha^4$. In QED perturbation theory there is the
electron vertex correction to the potential (4) which is defined by
the diagram in Fig.5 (a). In momentum representation the
corresponding operator of hyperfine interaction has the form:
\begin{equation}
\Delta V^{hfs}_{vertex}(k^2)=-\frac{8\alpha^2}{3m_em_\mu}
\left(\frac{{\mathstrut\bm\sigma}_e{\mathstrut\bm\sigma}_\mu}{4}\right)
\left[G_M^{(e)}(k^2)-1\right],
\end{equation}
where $G_M^{(e)}(k^2)$ is the electron magnetic form factor. We
extracted for the convenience the factor $\alpha/\pi$ from
$\left[G_M^{(e)}(k^2)-1\right]$. Usually used approximation for the
electron magnetic form factor $G_M^{(e)}(k^2)\approx
G_M^{(e)}(0)=1+\kappa_e$ is not quite correct in this task. Indeed,
characteristic momentum of the exchanged photon is $k\sim\alpha
M_\mu$. It is impossible to neglect it in the magnetic form factor
as compared with the electron mass $m_e$. So, we should use exact
one-loop expression for the magnetic form factor which was obtained
by many authors \cite{t4}. Let us note that the Dirac form factor of
the electron is dependent on the parameter of the infrared cutoff
$\lambda$. We take it in the form $\lambda=m_e\alpha$ using the
prescription $m_e\alpha^2\ll\lambda\ll m_e$ from Ref.\cite{HBES}.

\begin{figure}[htbp]
\centering
\includegraphics{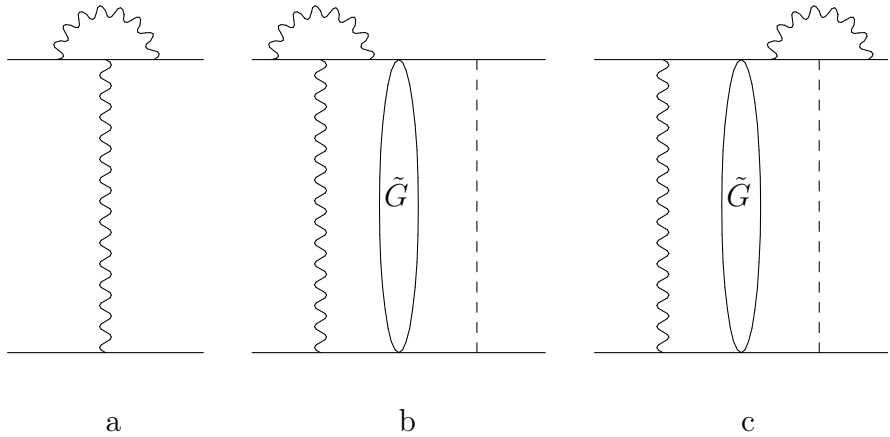}
\caption{The electron vertex corrections. The dashed line represents
the Coulomb photon. The wave line represents the hyperfine part of
the Breit potential. $\tilde G$ is the reduced Coulomb Green's
function.}
\end{figure}

Using the Fourier transform of the potential (52) and averaging the
obtained expression over wave functions (5) we represent the
electron vertex correction to the hyperfine splitting as follows:
\begin{equation}
\Delta\nu^{hfs}_{vertex}=\nu_F\frac{\alpha}{32\pi^2}\left(\frac{M_e}{M_\mu}\right)
\left(\frac{m_e}{\alpha M_\mu}\right)^3\int_0^\infty
k^2dk\left[G_M^{(e)}(k^2)-1\right]\times
\end{equation}
\begin{displaymath}
\times\left\{\left[1+\left(\frac{m_e}{4\alpha
M_\mu}\right)^2k^2\right]\left[\left(\frac{M_e}{2M_\mu}\right)^2+\left(\frac{m_e}{4\alpha
M_\mu}\right)^2k^2\right]^2\right\}^{-1}=4.214~MHz.
\end{displaymath}
Let us remark that the contribution (53) is of order $\alpha^5$.
Numerical value (53) is obtained after numerical integration with
the one-loop expression of the electron magnetic form factor
$G_M^{(e)}(k^2)$. If we use the value $G_M^{(e)}(k^2=0)$ then the
electron vertex correction is equal $5.244$ MHz. So, using the exact
expression of the electron form factors in the one-loop
approximation we observe the 1 MHz decrease of the vertex correction
to the hyperfine splitting from $1\gamma$ interaction. Taking the
expression (52) as an additional perturbation potential we have to
calculate its contribution to HFS in the second order perturbation
theory (see the diagram in Fig.5(b)). In this case the dashed line
represents the Coulomb Hamiltonian $\Delta H$ (3). Following the
method of the calculation formulated in previous section (see also
Refs.\cite{LM1,LM2}) we divide again total contribution from the
amplitude in Fig.5(b) into two parts which correspond to the muon
ground state $(n=0)$ and muon excited intermediate states $(n\not
=0)$. In this way the first contribution with $n=0$ takes the form:
\begin{equation}
\Delta
\nu^{hfs}_{vertex}(n=0)=\frac{8\alpha^2}{3\pi^2m_em_\mu}\int_0^\infty
k\left[G_M^{(e)}(k^2)-1\right]dk\int d{\bf x}_1\int d{\bf
x}_3\psi_{e0}({\bf x}_3)\times
\end{equation}
\begin{displaymath}
\times\Delta\tilde V_1(k,{\bf x}_3)G_e({\bf x}_1,{\bf x}_3)\Delta
V_2({\bf x}_1)\psi_{e0}({\bf x}_1),
\end{displaymath}
where $\Delta V_2({\bf x}_1)$ is defined by Eq.(38) and
\begin{equation}
\Delta \tilde V_1(k,{\bf x}_3)=\int d{\bf x}_4\psi_{\mu 0}({\bf
x}_4)\frac{\sin(k|{\bf x}_3-{\bf x}_4|)}{|{\bf x}_3-{\bf
x}_4|}\psi_{\mu 0}({\bf x}_4)=\frac{\sin\left(\frac{kx_3}{4\alpha
M_\mu}\right)}{x_3}\frac{1}{\left[1+\frac{k^2}{(4\alpha
M_\mu)^2}\right]^2}.
\end{equation}
Substituting the electron Green's function (19) in Eq.(54) we
transform desired relation to the integral form:
\begin{equation}
\Delta\nu^{hfs}_{vertex}(n=0)=\nu_F\frac{\alpha}{16\pi^2}\left(\frac{m_e}
{\alpha M_\mu}\right)^2\left(\frac{M_e}{M_\mu}\right)^2\int_0^\infty
\frac{k\left[G_M^{(e)}(k^2)-1\right]dk}{\left[1+\frac{m_e^2k^2}{(4\alpha
M_\mu)^2}\right]^2}\times
\end{equation}
\begin{displaymath}
\times\int_0^\infty x_3e^{-\frac{M_e}{2M_\mu}x_3}\sin\left(\frac{m_e
k}{4\alpha M_\mu}x_3\right)dx_3\int_0^\infty
x_1\left(1+\frac{x_1}{2}\right)e^{-x_1\left(1+\frac{M_e}{2M_\mu}\right)}dx_1
\times
\end{displaymath}
\begin{displaymath}
\left[\frac{2M_\mu}{M_ex_>}-\ln(\frac{M_e}{2M_\mu}x_<)-\ln
(\frac{M_e}{2M_\mu}x_>)+Ei(\frac{M_e}{2M_\mu}x_<)+
\frac{7}{2}-2C-\frac{M_e}{4M_\mu}(x_1+x_3)+
\frac{1-e^{\frac{M_e}{2M_\mu}x_<}}{\frac{M_e}{2M_\mu}x_<}\right]
\end{displaymath}
\begin{displaymath}
=-0.210~MHz.
\end{displaymath}
One integration over the coordinate $x_1$ is carried out
analytically and two other integrations are performed numerically.
Second part of the vertex contribution (Fig.5(b)) with $n\not =0$
can be reduced to the following form after several simplifications
which are discussed in section II (see also Refs.\cite{LM1,LM2}):
\begin{equation}
\Delta \nu^{hfs}_{vertex}(n\not
=0)=\nu_F\frac{8\alpha^4M_eM_\mu^3}{\pi^3}\int e^{-2\alpha M_\mu
x_2}d{\bf x}_2\int e^{-\alpha M_e x_3}d{\bf x}_3\int e^{-2\alpha
M_\mu x_4}d{\bf x}_4\times
\end{equation}
\begin{displaymath}
\times\int_0^\infty k\sin(k|{\bf x}_3-{\bf
x}_4|)\left(G_M^{(e)}(k^2)-1\right)\frac{|{\bf x}_3-{\bf
x}_2|}{|{\bf x}_3-{\bf x}_4|}\left[\delta({\bf x}_4-{\bf x}_2)-
\psi_{\mu 0}({\bf x}_4)\psi_{\mu 0}({\bf x}_2)\right].
\end{displaymath}
We divide expression (57) into two parts as provided by two terms in
the square brackets of (57). After that the integration (57) over
the coordinates ${\bf x}_1$, ${\bf x}_3$ is carried out
analytically. In the issue we obtain ($\gamma_1=M_e/4M_\mu$,
$\gamma_2=m_ek/4\alpha M_\mu$):
\begin{equation}
\Delta\nu^{hfs}_{1,vertex}(n\not=0)=\nu_F\frac{\alpha}{32\pi^2}\left(\frac{m_e}
{\alpha M_\mu}\right)^3\frac{M_e}{M_\mu}\int_0^\infty
k^2\left[G_M^{(e)}(k^2)-1\right]dk\times
\end{equation}
\begin{displaymath}
\times\left[\frac{4\gamma_1(\gamma_1^2-1)}{(1+\gamma_2^2)^3}-\frac{\gamma_1
(3+\gamma_1^2)}{(1+\gamma_2^2)^2}+\frac{4\gamma_1^2(\gamma_1^2-1)}{(\gamma_1^2+
\gamma_2^2)^3}+\frac{1+3\gamma_1^2}{(\gamma_1^2+\gamma_2^2)^2}\right]=2.516~MHz,
\end{displaymath}

\begin{equation}
\Delta\nu^{hfs}_{2,vertex}(n\not=0)=-\nu_F\frac{\alpha}{32\pi^2}\left(\frac{m_e}
{\alpha M_\mu}\right)^3\frac{M_e}{M_\mu}\int_0^\infty
k^2\left[G_M^{(e)}(k^2)-1\right]dk\times
\end{equation}
\begin{displaymath}
\times\frac{1}{(1+\gamma_2^2)^2}\left[\frac{2}{(\gamma_1^2+\gamma_2^2)}-\frac{(\gamma_1+1)}
{[(1+\gamma_1)^2+\gamma_2^2]^2}-\frac{2}{(\gamma_1+1)^2+
\gamma_2^2}-\frac{\gamma_2^2-3\gamma_1^2}{(\gamma_1^2+\gamma_2^2)^3}\right]=-0.831~MHz.
\end{displaymath}
It is necessary to emphasize that the theoretical error in the
contributions $\Delta\nu^{hfs}_{1,2,vertex}(n\not =0)$ is determined
by the factor $\sqrt{M_e/M_\mu}$ connected with the omitted terms of
the expansion similar to Eq.(25) (see also Refs.\cite{LM1,LM2}). It
can amount to $10\%$ of the results (58), (59) that is the value
near 0.2 MHz.

Until now we consider the electron vertex corrections connected with
the hyperfine part of the interaction Hamiltonian (4). But in the
second order perturbation theory we should analyze vertex
corrections to the Coulomb interactions of the electron and muon,
electron and nucleus. Then in the coordinate representation we have
the following potential:
\begin{equation}
\Delta V^C_{vertex, eN}(x_e)+\Delta
V^C_{vertex,e\mu}(x_{e\mu})=\frac{2\alpha^2}{\pi^2}\int_0^\infty
\frac{\left[G_E^{(e)}(k^2)-1\right]}{k}dk\left(\frac{\sin(kx_{e\mu})}{x_{e\mu}}-
2\frac{\sin(kx_e)}{x_e}\right),
\end{equation}
where we extract again the factor $\alpha/\pi$ from
$\left[G_E^{(e)}(k^2)-1\right]$. $G_E^{(e)}$ is the electron
electric form factor. One part of the contribution in Fig.5(c) is
specified by the electron-muon intermediate states in which the muon
is in the ground state $n=0$. This correction is determined by both
terms in the round brackets of Eq.(60) and can be presented as
follows:
\begin{equation}
\Delta\nu^{hfs}_{C,vertex}(n=0)=\nu_F\frac{\alpha}{\pi^2}
\left(\frac{M_e}{M_\mu}\right)^2\int_0^\infty x_3^2
e^{-x_3\left(1+\frac{M_e}{2M_\mu}\right)}dx_3\times
\end{equation}
\begin{displaymath}
\times\int_0^\infty x_1e^{-\frac{M_e}{2M_\mu}x_1}dx_1 \int_0^\infty
\frac{\left[G_E^{(e)}(k^2)-1\right]dk}{k}\sin\left(\frac{m_ek}{4\alpha
M_\mu}x_1\right)\left\{1-\frac{1}{2\left[\frac{m_e^2k^2}{(4\alpha
M_\mu)^2}+1 \right]^2}\right\}\times
\end{displaymath}
\begin{displaymath}
\left[\frac{2M_\mu}{M_ex_>}-\ln(\frac{M_e}{2M_\mu}x_<)-\ln
(\frac{M_e}{2M_\mu}x_>)+Ei(\frac{M_e}{2M_\mu}x_<)+
\frac{7}{2}-2C-\frac{M_e}{4M_\mu}(x_1+x_3)+
\frac{1-e^{\frac{M_e}{2M_\mu}x_<}}{\frac{M_e}{2M_\mu}x_<}\right]
\end{displaymath}
\begin{displaymath}
=-1.321~MHz.
\end{displaymath}
The index "C" means that the vertex correction to the Coulomb part
of the Hamiltonian is considered. Excited states of the muon $(n\not
=0)$ contribute to another part of the matrix element (Fig.5(c)).
Changing the Coulomb Green's function of the electron by free
Green's function (see discussion in section II) we can make the
coordinate integration and express the correction to HFS as
one-dimensional integral:
\begin{equation}
\Delta\nu^{hfs}_{C,vertex}(n\not =
0)=-\nu_F\frac{8\alpha}{\pi^2}\frac{M_e}{M_\mu}\left(\frac{\alpha
M_\mu}{m_e}\right)\int_0^\infty
\frac{\left[G_E^{(e)}(k^2)-1\right]dk}{k^2}
\left\{1-\frac{1}{\left[1+\frac{m_e^2k^2}{(4\alpha M_\mu)^2}
\right]^4}\right\}=
\end{equation}
\begin{displaymath}
=1.770~MHz.
\end{displaymath}
The electron vertex corrections investigated in this section have
the order $\alpha^5$ in the hyperfine interval. Summary value of all
obtained contributions (53), (56), (58), (59), (61), (62) is equal
to 6.138 MHz. It differs by a significant value 0.894 MHz from the
result 5.244 MHz which was used previously by many authors for the
estimation of the electron anomalous magnetic moment contribution.

\section{Conclusions}

In the present study, we have performed the analytical and numerical
calculation of several important contributions to the hyperfine
splitting of the ground state in muonic helium atom connected with
the vacuum polarization, the nuclear structure effects and the
electron vertex corrections. To solve this task we use the method of
the perturbation theory which was formulated previously for the
description of the muonic helium hyperfine splitting in
Refs.\cite{LM1,LM2}. We have considered corrections of order
$\alpha^5$ of the electron vacuum polarization and electromagnetic
form factors and the nuclear structure effects of order $\alpha^6$.
The numerical values of the corresponding contributions are
displayed in Table I. We present in Table I the references to the
calculations of other corrections which are not considered here. The
relativistic correction was obtained in Ref.\cite{HH}, the vertex
correction was calculated in the case of hydrogenic atoms in
Refs.\cite{EGS,BE,KP,KKS}. Basic contributions to the hyperfine
splitting obtained by Lakdawala and Mohr are also included in Table
I because our calculation is closely related to their approach.

Let us list a number of features of the calculation.

1. For muonic helium atom, the vacuum polarization effects are
important and give rise to the modification of the two-particle
interaction potential which provides the
$\alpha^5\frac{M_e}{M_\mu}$-order corrections to the hyperfine
structure. The next to leading order vacuum polarization corrections
(two-loop vacuum polarization) are negligible.

2. The electron vertex corrections should be considered with the
exact account of the one-loop electromagnetic form factors of the
electron because the characteristic momentum incoming in the
electron vertex operator is of order of the electron mass.

3. In the $\alpha^6$-order the nuclear structure corrections to the
ground state hyperfine splitting are expressed in terms of the
charge radius of $\alpha$-particle.

4. Analyzing the one-loop electron vacuum polarization and vertex
effects and the nuclear structure contributions in each order in
$\alpha$, we have taken into account recoil terms proportional to
the ratio of the electron and muon masses.

The resulting numerical value 4465.526 MHz of the ground state
hyperfine splitting in muonic helium is presented in Table I. It is
sufficiently close both to the experimental result (1) and the
earlier performed calculations by the perturbation theory,
variational approach and the Born-Oppenheimer theory: $4464.3\pm
1.8$ MHz \cite{LM2}, $4465.0\pm 0.3$ MHz \cite{HH1}, 4462.9 MHz
\cite{Amusia}, $4450.4\pm 0.4$ MHz \cite{RD}, $4459.9$ MHz
\cite{RD1}, $4464.87\pm 0.05$ MHz \cite{Chen1}. The estimation of
the theoretical uncertainty can be done in terms of the Fermi energy
$\nu_F$  and small parameters $\alpha$ and the ratio of the particle
masses. On our opinion there exist several main sources of the
theoretical errors. First of all, as we mentioned above
comprehensive analytical and numerical calculation of recoil
corrections of orders $\alpha^4\frac{M_e}{M_\mu}$,
$\alpha^4\frac{M^2_e}{M^2_\mu}$,
$\alpha^4\frac{M^2_e}{M^2_\mu}\ln(M_\mu/M_e)$ was carried out by
Lakdawala and Mohr in the second order PT in Refs.\cite{LM1,LM2}.
The error of their calculation connected with the correction
$\nu_F\frac{M_e^2}{M_\mu^2}\ln\frac{M_\mu}{M_e}$ consists 0.6 MHz.
The second source of the error is related to contributions of order
$\alpha^2 \nu_F\approx 0.2$ MHz which appear both from QED
amplitudes and in higher orders of the perturbation theory. Another
part of the theoretical error is determined by the two-photon
three-body exchange amplitudes mentioned above. They are of the
fifth order over $\alpha$ and contain the recoil parameter
$(m_e/m_\alpha)\ln(m_e/m_\alpha)$, so that their possible numerical
value can be equal $\pm 0.05$ MHz. Finally, a part of theoretical
error is connected with our calculation of the electron vertex
corrections of order $\alpha^5$ in section IV. It consists at least
0.2 MHz (see the discussion after Eq.(59)) We neglect also the
electron vertex contributions of order $\nu_F\alpha M_e/M_\mu\approx
0.2$ MHz which appear in higher orders of the perturbation theory.
Thereby, the total theoretical uncertainty is not exceeded $\pm 0.7$
MHz. The existing difference between the obtained theoretical result
and experimental value of the hyperfine splitting (1) equal to 0.522
MHz lies in the range of total error. Theoretical error which
remains sufficiently large in the comparison with the experimental
uncertainty, initiates further theoretical investigation of the
higher order contributions including more careful construction of
the three-particle interaction operator connected with the
multiphoton exchanges.

\begin{acknowledgments}

We thank Organizing Committee of the Conference of Nuclear Physics
Department RAS  in ITEP (Moscow) for the financial support. The
final part of the work was carried out in the Humboldt University in
Berlin. One of the authors (A.P.M.) is grateful colleagues from
Institute of Physics for warm hospitality. This work was supported
in part by the Russian Foundation for Basic Research (Grant
No.06-02-16821).
\end{acknowledgments}

\end{document}